\documentclass[12pt,preprint]{aastex}

\def\etal{{\it et al. }}
\def\kms{km~s$^{-1}$}
\def\sun{$_{\mathrm \odot}$}
\def\ergs{erg~s$^{-1}$}
\newcommand{\bd}{\begin{displaymath}}
\newcommand{\ed}{\end{displaymath}}
\usepackage[usenames,dvips]{color}

\slugcomment{submitted to ApJ,\today}
\shorttitle{}
\shortauthors{Wang et al.}

\begin{document}
\title{Metal-Enriched Outflows in the Ultra-Luminous infrared Quasar Q1321+058}
\author{Tinggui Wang$^{1,2}$, Hongyan Zhou$^{3,1,2}$, Weimin Yuan$^4$, HongLin Lu$^2$,
        Xiaobo Dong$^{1,2}$ and Hongguang Shan$^{4}$}
\affil{CAS Key Laboratory for Research in Galaxies and Cosmology, University of Science and Technology of China, Hefei, Anhui, 230026, P.R.China}
\affil{Center for Astrophysics, University of Science and Technology
       of China, Hefei, Anhui, 230026, P.R. China}
\affil{Department of Astronomy, University of Florida, Gainesville,
FL 32611} 
\affil{ Yunnan Astronomical Observatory, 
National Astronomical Observatories, Chinese Academy of
Sciences, P.O. Box 110, Kunming, Yunnan 650011, China}

\begin{abstract}

Quasar outflows may play important role in the evolution of its host galaxy 
and central black hole, and are most often studied in absorption lines. In 
this paper, we present a detailed study of multiple outflows in the obscured 
ultra-luminous infrared quasar Q1321+058. The outflows reveal themselves in 
the complex optical and UV emission line spectrum, with a broad component  
blueshifted by 1650\kms and a narrow component by 360\kms, respectively. 
The higher velocity component shows ever strong \ion{N}{3}] 
(\ion{N}{3}]/\ion{C}{3}]=3.8$\pm$ 0.3 and \ion{N}{3}]/\ion{C}{4}=0.53) 
and strong \ion{Si}{3}] (\ion{Si}{3}] /\ion{C}{3}]$\simeq$ 1), in addition 
to strong [\ion{O}{3}]$\lambda$5007 and [\ion{Ne}{3}]$\lambda$3869 emission. 
A comparison of these line ratios with photoionization models suggests an 
overabundance of N and Si relative to C. The abundance pattern is consistent 
a fast chemical enriching process associated with a recent starburst, 
trigerred by a recent galaxy merger. The outflow extends to several tens 
to hundred parsecs from the quasar, and covers only a very small sky. We 
find that the outflow with line emitting gas is energetically 
insufficient to remove the ISM of the host galaxy, but total kinetic energy 
may be much larger than suggested by the emission lines. The velocity range 
and the column density suggest that the outflow might be part of the low 
ionization broad absorption line region as seen in a small class of quasars. 

The optical and UV continuum is starlight-dominated and can be modeled with 
a young-aged (1 Myr) plus an intermediate-aged ($\sim 0.5-1$ Gyr) stellar 
populations, suggesting a fast building of the stellar mass in the host 
galaxy, consistent with the starburst-type metal abundances inferred from 
the high velocity outflow spectrum. The broad band spectral energy 
distribution shows that it is an obscured quasar with its bulk emission 
in the middle infrared. The star formation rate, independently estimated 
from UV, far-infrared, and emission line luminosity, is much lower than 
that is required for the co-evolution of the black hole and its host 
spheroid. 
\end{abstract}


\section{Introduction}

A consensus is now forming that quasar activities are triggered through 
mergers of gas-rich massive galaxies. The evidence accumulated in the past 
decade includes the presence of tidal tails in their host galaxies 
\citep[e.g.][]{kir99,hut99}, and signatures of a recent starburst  
in the off-nuclear spectrum \citep[e.g.][]{cs00a,cs00b} and in the 
composite spectrum of less luminous AGN \citep[e.g.][]{hec04,van06}. 
The merger of gas-rich galaxies also leads to starburst in the nuclear 
as well as outskirt regions. As such, most of these systems appear as 
luminous or ultra-luminous infrared galaxies ($L_{ir}>10^{12}L$\sun, 
ULIRGs). The connection between ULIRGs and quasars has been widely 
discussed, and there is evidence for an evolution sequence between 
the two \citep[e.g.][]{san96,zheng99}.

Vigorous starburst and nuclear activity will result in a strong 
negative feedback to the host galaxy. Massive winds have been observed in
starburst galaxies, ULIRG and AGN with mass outflow rates from 10 to
1000 M\sun~yr$^{-1}$ \citep[e.g.][]{rup05,lipari05}. It was proposed
that such a feedback will quench gas supply to the AGN and the
star-formation region, and result in a sudden halt of the
star-formation process in the galaxies \citep[e.g.][]{dim05}. The
passively evolved galaxies will first appear as E+A in the
subsequent $\sim$ 1 Gyr, and then as red galaxies. The latter can
explain the red colors of massive early type galaxies \citep{spr05}.
If quasar activity is going-on for sometime  after the
star-formation essentially ceased, the object will show
characteristics of Q+A. Q+A's have been detected now with a large
number in the Sloan Digital Sky Survey \citep[SDSS]{york00}. 
Most of these systems show weak [\ion{O}{2}] emission, likely from 
the Narrow Line Region (NLR), indicating that major star-forming 
activity has ceased\citep{zhou05}, or, at least, is greatly 
suppressed. Indeed, Ho (2005) found that the ratio of SFR to mass 
accretion rate in PG quasars is well below that required for 
sustaining the $M_{\mathrm BH}-\sigma$
relation for nearby spheroid galaxies. It was further suggested
that this is due to that star formation is suppressed rather than 
that cold gas is exhausted. It should be noted that the author used the 
[\ion{O}{2}] emission-line luminosity to infer SFR,  for which the reddening 
correction is largely uncertain, however.

Gas outflows are thought to play an important role in the galactic scale 
feedback. Outflows on scales of tens to hundreds parsecs have been observed 
in nearby Seyfert galaxies and radio galaxies with velocities of a few
hundreds to thousand \kms~\citep[e.g.][]{gab05,das05,ruiz05}. However, 
modeling the absorption and emission lines for well-studied nearby Seyfert 
galaxies suggested a relative low mass loss rate and small kinetic power 
in these outflows (Crenshaw, Kraemer \& George 2003). Even higher velocity outflows, 
from a few $10^3$ to $10^4$ \kms, are detected in broad absorption line 
(BAL) quasars \citep[e.g.][]{wey91}, but the scale and the total mass in 
BAL outflows are still not clear,  neither the relation between the 
BAL region (BALR) and the narrow/broad emission-line region (N/BELR). 

The recent intensive starburst in the circum-nuclear region has also 
significant implications for the gas phase metallicity. Massive stars 
evolve very fast and produce a chemical pollution in the circum-nuclear ISM 
primarily with $\alpha$-elements via S\ion{N}{2}, while less massive stars evolve 
much slowly and make their chemical contribution of iron peaked elements, 
mainly through SNIa, with a delay of at least 1 Gyr. The detail abundance 
pattern depends on both the history of star-formation and the stellar 
initial mass function. Supersolar metallicity has been suggested 
by analyzing the broad line and broad absorption line spectra 
(e.g., Hamann 1997; Baldwin et al. 2003) 
, which are produced in a relative 
small region enclosed the active nucleus. 
However, it is still not clear whether this reflects an over-all high 
metallicity or only these elements concerned.  

In this paper, we study in detail the optical and ultraviolet (UV) 
spectra and the broad band properties of the ULIRG Q 1321+058. 
This object was classified as a quasar based on a low resolution 
spectrum taken in the optical identification processes of HEAO~1-A2 
(The High Energy Astrophysics Observatory-1) X-ray sources \citep{rem93}. 
However, it was not detected by XMM-Newton \citep{bia05}, which 
led these authors to propose that it is a Compton-thick, type II object. 
The optical image taken with Hubble Space Telescope (HST) showed a strong 
disturbed morphology possibly with a double nucleus of separation less 
than 1 kpc, suggesting a recent merger of two gas-rich galaxies 
\citep{boyce96}. Darling \& Giovanelli (2002) detected an OH Megmaser 
with a multi-peaked profile, which they interpreted as possible multiple 
masser nuclei. 
Here we show that its broad band properties are consistent with an 
obscured quasar. The optical and the UV continuum is likely dominated 
by starlight, that is also typical for a type II object. 
Extreme velocity outflows (EVOFs) were noticed previously in the 
[\ion{O}{3}] and H$\beta$ emission-lines \cite{lipari03}. Taking the 
advantage of the broad wavelength coverage of the SDSS spectrograph, 
we have identified the EVOFs in H$\alpha$ and [\ion{Ne}{3}]$\lambda$3869. We 
find that the EVOFs also appear in the UV lines in the HST (Hubble 
Space Telescope) FOS spectrum, such as \ion{C}{4}$\lambda$1549, \ion{N}{3}]$\lambda$1750, 
\ion{Al}{3}]$\lambda$1860, \ion{Si}{3}]$\lambda$1892, and \ion{C}{3}]$\lambda$1909. 
The line ratios imply a dense line emitting gas with a fast metal 
enriched history. We find with interest that the velocity range and the 
column density of the AGN-driven outflows are comparable to those in 
low-ionization BALR
(LoBALR), implying that the object would appear as a LoBAL quasar if 
our line-of-sight passing through the outflows.
These unusual properties of Q 1321+058 provides us with an ideal
laboratory to study the impact of AGN and starburst feed-backs to
galaxy formation and evolution. 
Throughout this paper, we
assume a $\Lambda$-CDM cosmology with
$H_0=72~$\kms~Mpc$^{-1}$, $\Omega_\Lambda=0.7$ and
$\Omega_m=0.3$.

\section{Modeling Continuum and Emission Line}

The SDSS spectrum used in this work was extracted from the
SDSS archive\footnote{http://www.sdss.org/.}. The UV spectrum was
observed with the high-resolution gratings of the Faint Object
Spectrograph (FOS) on board HST (Bechtold et al. 2002). 
We retrieved the calibrated 1-D spectrum from the HST archive 
using the on-flight calibration. The spectra of two exposures 
are combined to remove cosmic rays. 
Both optical and UV spectra are wavelength- and
flux-calibrated. 
The spectra are corrected for 
Galactic reddening of E(B-V)=0.031 \citep{sch98} and
are shifted into the rest frame of the object using a 
redshift of $z=0.20467$ as
measured from optical emission lines (see below).

\subsection{ Continuum}
\begin{figure}
\plotone{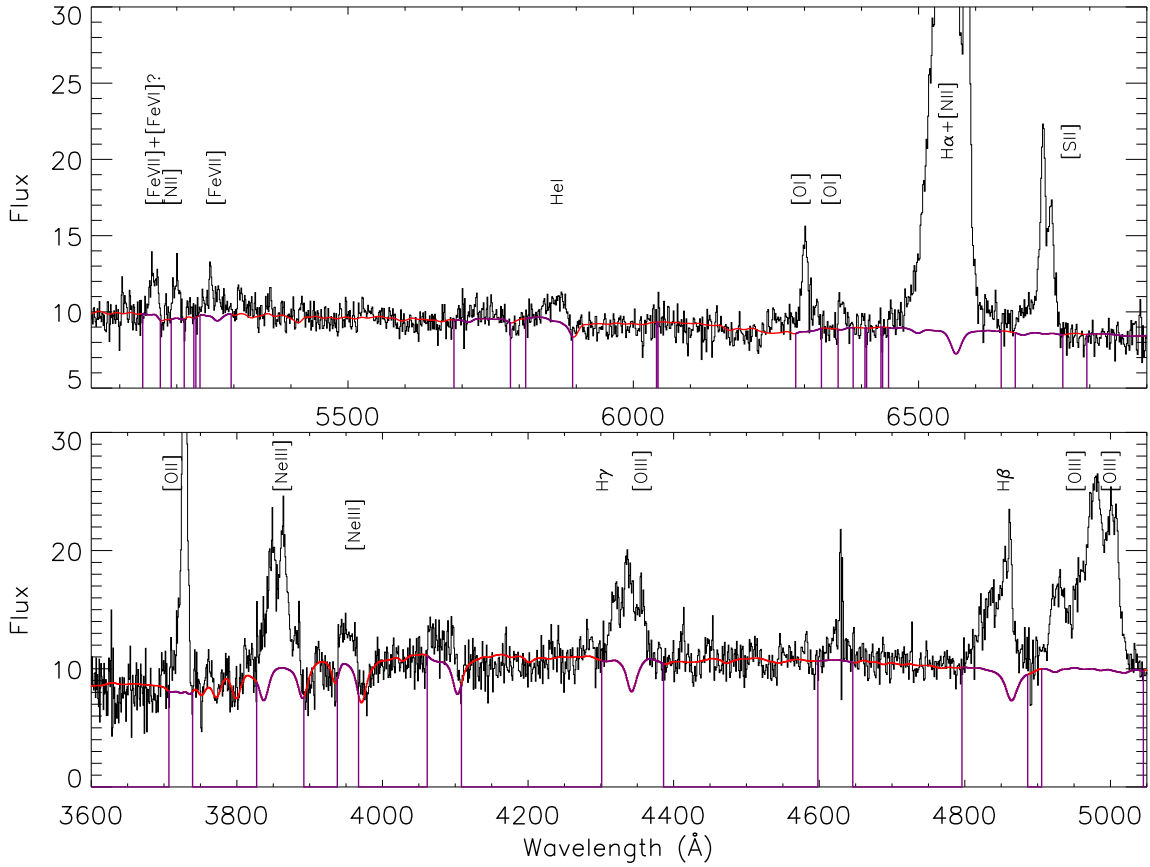}
\caption{Optical (SDSS) spectrum of Q1321+058 and the best-fit star
light model (see text for details). Prominent emission lines are 
labeled. Masked regions during the stellar light fit are marked in 
purple color. 
 }
\label{fig1}
\end{figure}

The optical continuum shows prominent high order Balmer absorption 
lines, thus is likely dominated by starlight (Fig \ref{fig1}). We 
model the continuum with the templates derived using the ICA method 
(refer to Lu et al. (2006) for detail) assuming no contribution 
from the AGN:
\begin{equation}
f(\lambda)=A(\lambda)\sum a_i IC_i(\lambda)
\end{equation}
where $A(\lambda)$ is the dust extinction factor, $IC_i$ the $i$th
independent component. Note that both $IC_i$ and $a_i$ are positive. 
The intrinsic reddening is taken as a free parameter and the extinction 
curve of Calzetti et al. (2000) for starburst galaxies is adopted. The 
templates are broadened and shifted 
, up to 1000 km~s$^{-1}$ in velocity space, to match the observed 
spectrum. With the redshift noted above, we find that 
$\Delta v=-40\pm 30$~\kms. Emission line regions were masked during the 
fit. The best-fit starlight continuum is shown in Fig \ref{fig1}. The 
presence of Balmer absorption lines and lack of an apparent 4000\AA~ 
suggest a recent/on-going starburst. An effective reddening of 
starlight is yielded to be $E(B-V)=0.65\pm0.05$ for the Calzetti 
\etal's extinction curve, which is among the most dusty galaxies 
in the SDSS galaxy sample\citep{lu06}.

Considering the relative low S/N ratio of the HST-FOS spectrum, we fit 
the UV continuum with either a reddened simple stellar population (SSP) 
or a reddened power-law. Prominent emission lines are modeled with 
Gaussians in this fit with their widths and centroids as free parameters; 
these include \ion{C}{4}]$\lambda$1549, \ion{Si}{4}+\ion{O}{4}]$\lambda$1400, \ion{He}{2}$\lambda$1640, 
\ion{O}{3}]$\lambda$1663, \ion{N}{3}]$\lambda$1750, \ion{Al}{3}$\lambda$1860, \ion{N}{4}$\lambda$1486
\ion{Si}{3}]$\lambda$1892, and \ion{C}{3}]$\lambda$1909. Weak emission lines 
detected only marginally, such as \ion{Si}{2}$\lambda \lambda$1808,1817, 
are not considered. The prominent emission line centered at 
$\lambda 1782.4$ rest frame, which can be identified as blue shifted 
\ion{Fe}{2} 191 (wavelength 1787\AA), is modeled with a Gaussian. In the worst 
case, wavelength zero point of FOS can be displaced up to 
250 km~s$^{-1}$ (Keyes, HST Handbook), we will keep this in mind in the 
following analysis of the FOS data.

\begin{figure}
\plotone{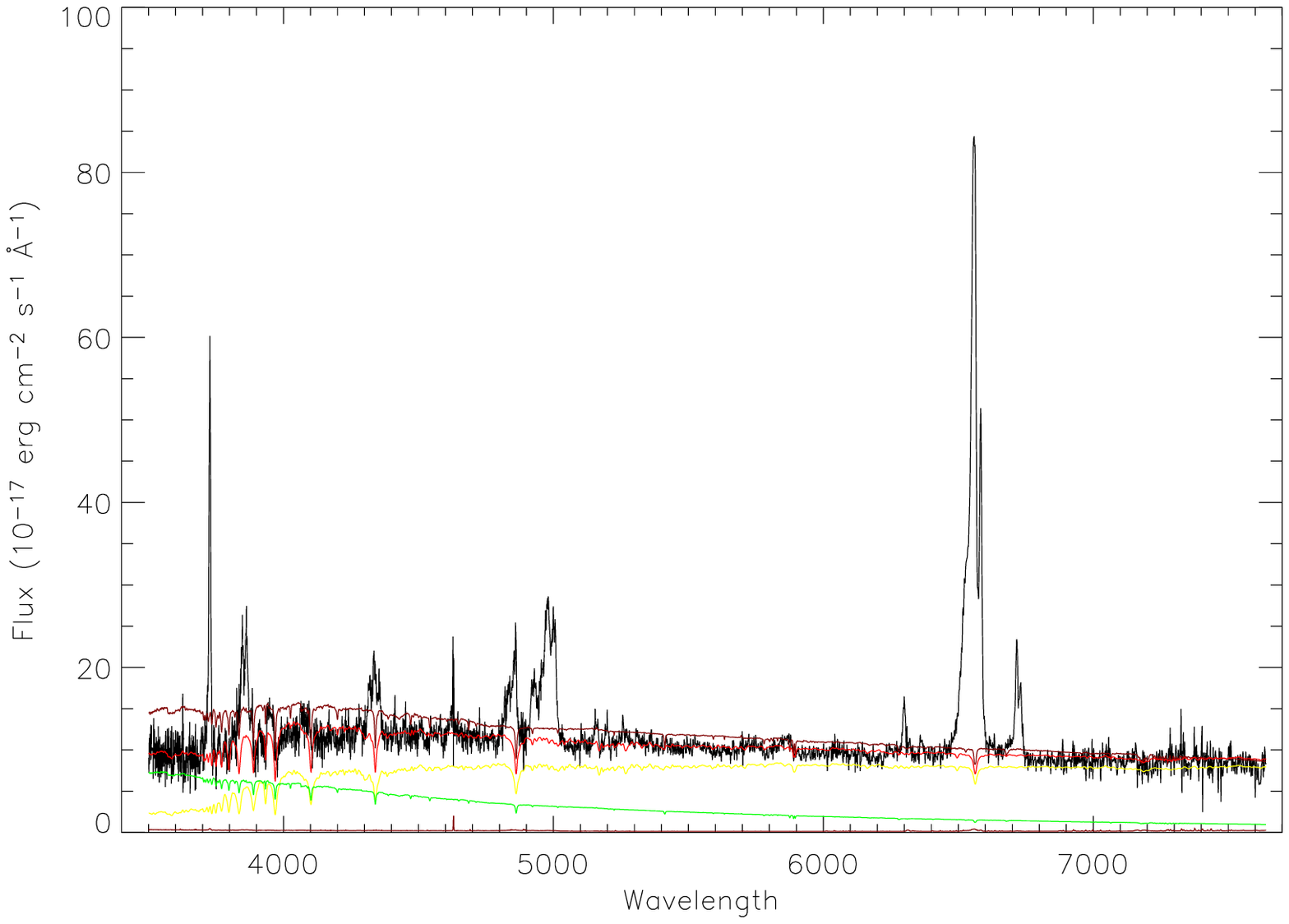}
\plotone{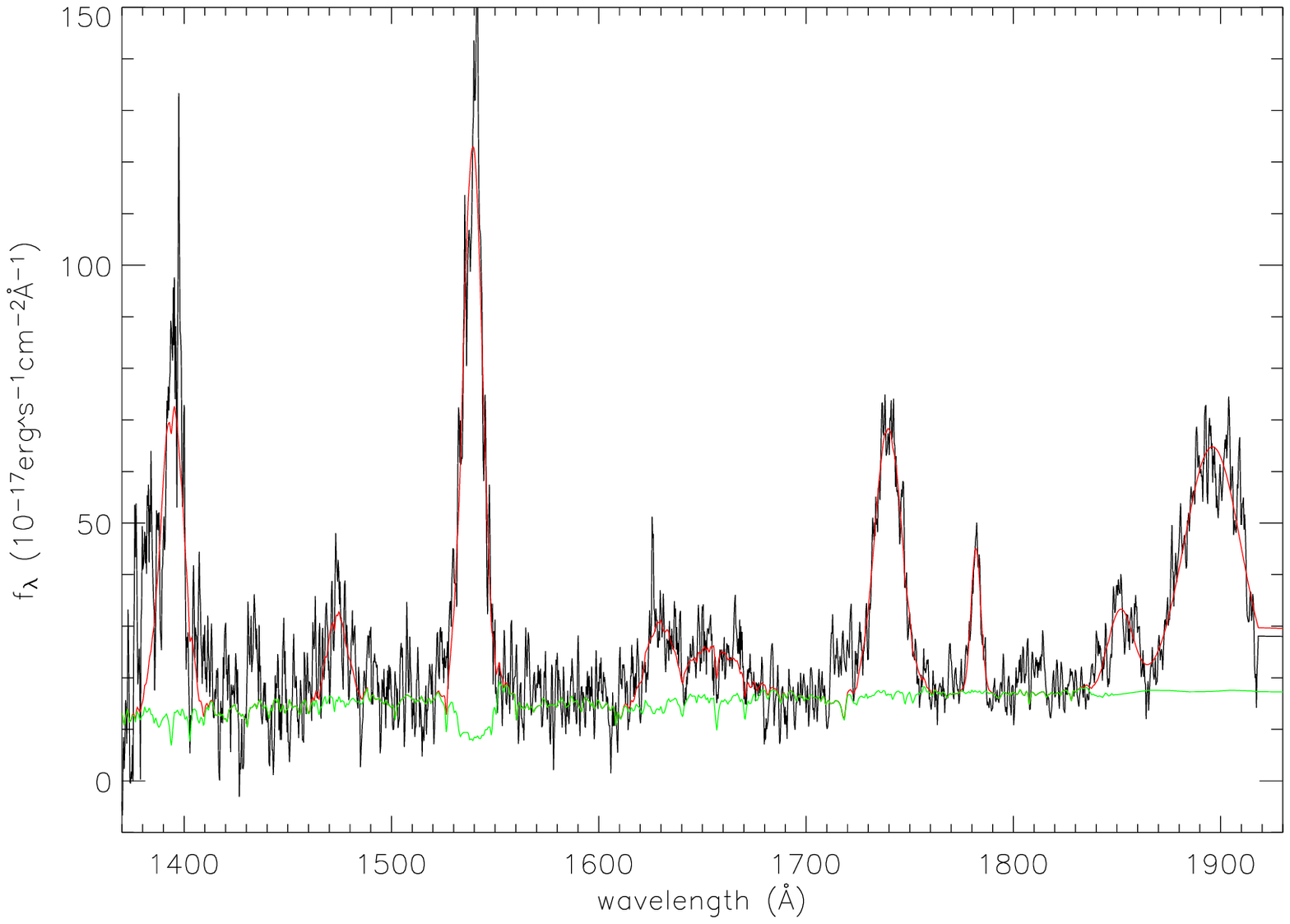}
\caption{Upper panel: The two-stellar population model for the optical
spectrum. The UV model with the SMC (Calzetti et al's) extinction curve is
shown in green (brown) color, and the intermediate 
age stellar population in yellow, and the final model in red. 
Lower panel: The UV spectrum
and the best-fit reddened single stellar population model (the SMC
extinction curve; Green curves). 
The fit to the UV continuum and emission lines is shown in red.
}
\label{fig2}
\end{figure}

The SSP spectra $f_{SB99}(\lambda)$ was taken from starburst99 models 
\citep{lei99,vaz05}. The high resolution SSP models are only available 
for the solar abundance with a wavelength coverage of $1200-1870~\AA$, 
and with ages from 1 to 20 Myr for either an instantaneous burst or a 
continuous star formation. To extend to the full wavelength range of 
the FOS spectrum, we use the interpolated low resolution model of 
Starburst99 for $\lambda>$1870\AA. This is reasonable since the 
extrapolated wavelength range is rather small ($\Delta \lambda=50~\AA$) 
and free from strong stellar absorption lines. The UV continuum can be 
reasonably well fitted with this simple model (Fig \ref{fig2}). We use 
both SMC and Calzetti's extinction curves. But the best fitted model 
with Calzetti's extinction curve predicts an 
optical flux well above the SDSS spectrum, thus will not be discussed 
further. The best fitted model converges to a stellar
population (1~Myr) and a moderate reddening $E(B-V)\sim 0.26\pm 0.03$
for the SMC extinction curve. The SFR rate is 450$\pm$50 M\sun~yr$^{-1}$
for instantaneous burst and 430$\pm$40 M\sun~yr$^{-1}$ for continuous
formation models after scaled to a lower stellar mass cutoff of
0.1~M\sun. For instantaneous model 2~Myr population yields significant
worse fit ($\Delta\chi^2=10$), however, for continuous formation model,
2~Myr population also yields reasonable fit ($\Delta\chi^2=3$) with a
factor of two smaller SFR 230$\pm$20~M\sun~yr$^{-1}$. Note that only 
statistical uncertainty is quoted here, while much large errors may be 
introduced by our assumptions, such as uniform dust reddening and ignoring 
intermediate age stellar populations. Thus the SFR value must be interpreted 
with care. 

The optical spectrum, obtained by subtracting the best UV model, can 
be fitted with a reddened stellar population of an intermediate age 
around 0.90 Gyr (Fig \ref{fig2}), with a reddening of $E(B-V)\approx 
0.30$ using Calzetti et al.'s extinction curve. The derived stellar 
mass is 9$\times10^{10}$ M\sun. Using the SMC extinction curve, we 
obtain a similar reddening ($E(B-V)=0.40$) and a slightly younger 
stellar population (0.64 Gyr) with a somewhat smaller total stellar mass 
(13\%~less). With those exercises, albeit with some uncertainties, 
we conclude that a few $\sim 10^{10}$ M\sun stellar mass has been 
built within the last Gyr in the host of Q 132+058. Note the older 
stellar population has a negligible contribution to the UV flux. 

The power-law model yields a fit to UV spectrum as good as the SSP 
model. Therefore, we 
cannot distinguish two models from the UV fit alone. If both power-law 
index and reddening vary freely, they are poorly constrained. For 
an unreddened power-law, a slope of $\beta=2.0\pm0.2$ ($f_\lambda\propto
\lambda^\beta$) is obtained. Fixed the slope to typical quasar value 
$\beta=-1.7$, a reddening of $E(B-V)=0.28\pm 0.02$ is yielded for 
SMC-type grain. However, extrapolating these models over-predicts 
severely the optical flux below 4000\AA. Only when $\beta\leq -6.0$, 
the model flux is consistent with the optical flux with a large intrinsic 
reddening $E(B-V)\geq 0.605$. Not only such a flat UV spectrum has never been 
observed in quasars, but also the reddened corrected UV flux around 1400\AA~ 
is a factor of 3600 times the observed value, which gives an UV luminosity 
well larger than the total infrared luminosity estimated
in \S \ref{activity}. Thus we believe that a single power-law model is 
not realistic.   

\subsection{Emission lines}

\begin{figure}
\plotone{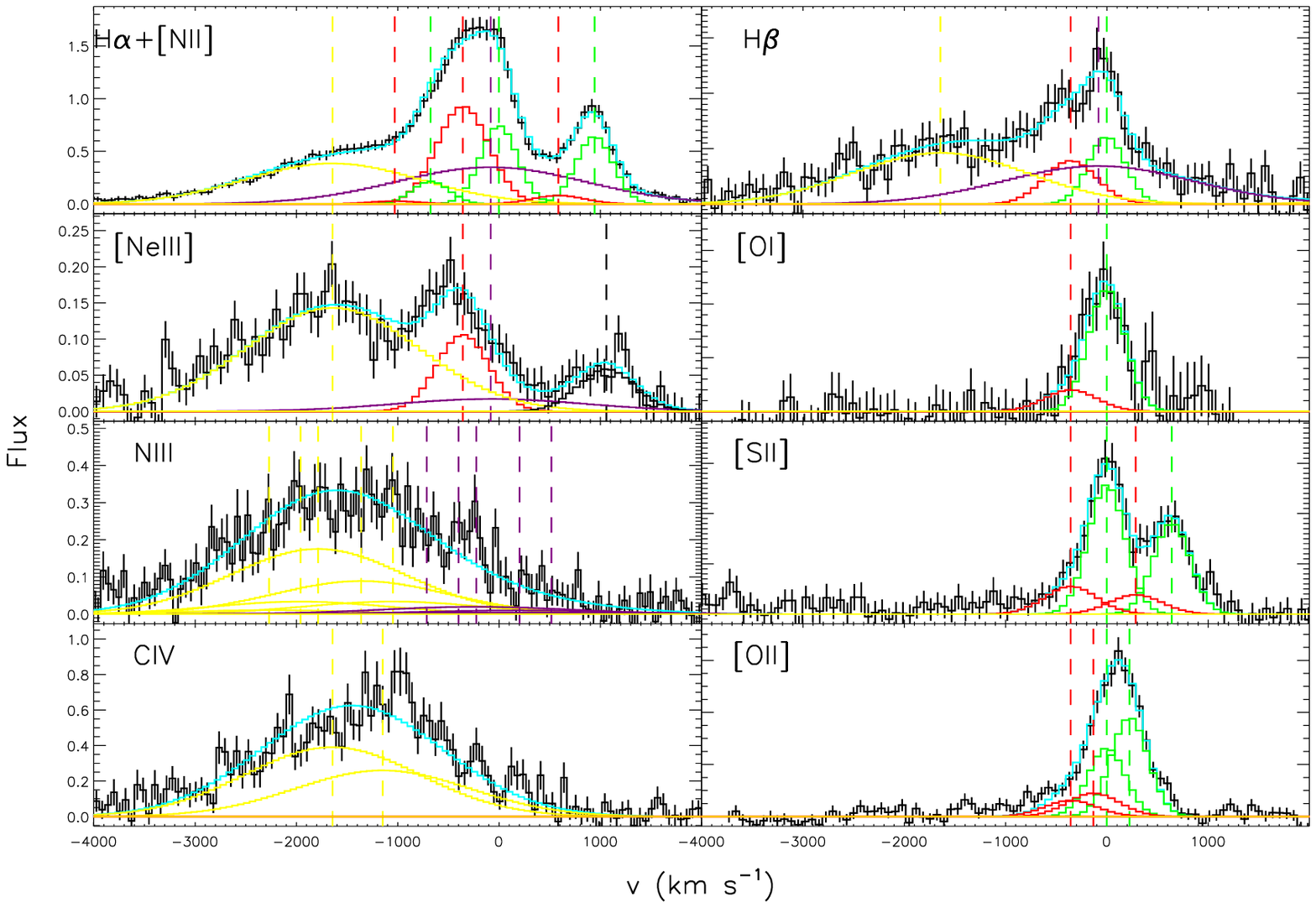}
\plotone{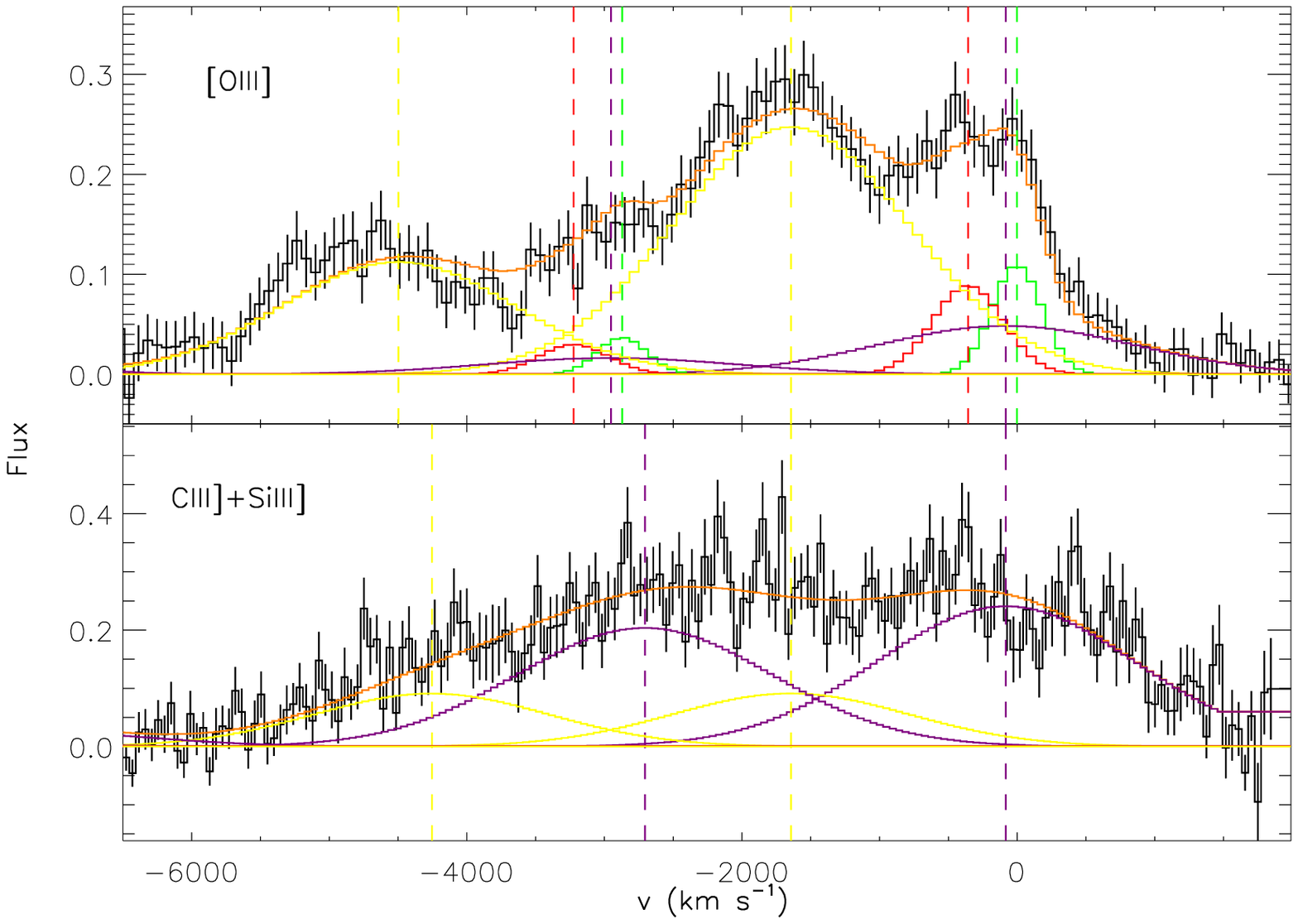}
\caption{Continuum subtracted emission line profiles and the best-fit 
four component models, green for C1, red for C2, purple for C3, and yellow 
for C4 (see text for details).
}
\label{fig3}
\end{figure}

We obtain the emission-line spectrum by subtracting the model continuum. 
The emission-line profiles are rather complex and show several distinct 
emission-line components (see Fig \ref{fig3}): 
\begin{description}
\item[C1] The low-ionization forbidden lines, 
      [\ion{O}{2}]$\lambda\lambda$3726,3729, 
      [\ion{O}{1}]$\lambda$6302, [\ion{S}{2}]$\lambda\lambda$6717,6732 and 
      [\ion{N}{2}]$\lambda$6584 
      display a single-peaked profile at a redshift of z=0.20467, which can be 
      taken as the systematic velocity. There is a corresponding peak in 
      H$\beta$, [\ion{O}{3}]$\lambda$5007. However, this component is absent in 
      the ultraviolet lines. 

\item[C4] The isolated UV lines \ion{C}{4}$\lambda$1549 and 
      \ion{N}{3}]$\lambda$1750 show 
      a single-peaked profile, which can be well fitted with a Gaussian 
      blueshifted by $\sim$ 1,800 \kms~relative to the systematic redshift. 
      This component is also evident in [\ion{Ne}{3}]$\lambda$3869, and 
      [\ion{O}{3}]$\lambda$4959 and in [\ion{O}{3}]$\lambda$5007, 
      which is blended 
      with the C3 of [\ion{O}{3}]$\lambda$4959. 
      The extended blue wing of Balmer 
      lines, and \ion{C}{3}], \ion{Si}{3}] can be attributed to 
      this component, too.

\item[C2] [\ion{Ne}{3}] shows an additional component centered at the velocity 
      around -500\kms. A peak at this velocity is also evident in 
      [\ion{O}{3}]$\lambda$5007, H$\beta$ line. It is likely the weak blue 
      wing in the [\ion{O}{2}]$\lambda$3727, [\ion{S}{2}] and [\ion{O}{1}] 
      is due to this component, too. A comparison of H$\alpha$ line profile 
      with those of low ionization lines such as [\ion{S}{2}], [\ion{O}{2}], 
      [\ion{O}{1}] and [\ion{N}{2}] also suggests such a component in 
      H$\alpha$. 

\item[C3] Both \ion{C}{3}], H$\beta$ and [\ion{O}{3}] show extended red wing up 
      to a velocity of $\sim$1,500\kms, which is not present in [\ion{O}{2}], 
      [\ion{S}{2}], \ion{C}{4} and \ion{N}{3}. Note that the redside of 
      [\ion{Ne}{3}]$\lambda$3869 is affected by both \ion{He}{1} and 
      [\ion{Fe}{7}]$\lambda$3890, thus it is not 
      clear if this component is also present in [\ion{Ne}{3}].    

\end{description}

C1 and C4 are well defined from the isolated lines such as \ion{C}{4}, 
\ion{N}{3}], [\ion{S}{2}], [\ion{O}{1}], and [\ion{O}{2}]. One gaussian 
can adequately describe these components. C2 and C3 are less cleanly 
defined due to their blending with other components. It is interesting 
to note that the velocity separation between C1 and C2 is very close 
to the separation (490\kms) of two peaks detected in the OH maser 
profile (Darling \& Giovanelli 2002). In the following fit, we will 
use a gaussian for all components. The good fit to the 
\ion{C}{3}]+\ion{Si}{3}] blending with only C3 and 
C4 components indicates that C3 can be approximately described with a 
gaussian. While a gaussian C2 component yields also reasonable fit to 
the optical lines suggests that such approximation is acceptable. 

Note in passing, the prominent emission line at 1782.4$\pm$0.2\AA~ has  
a width of $\sigma = 387\pm 34$~\kms~ and flux of (189$\pm$14)$\times 
10^{-17}$~ erg~cm$^{-2}$~s$^{-1}$ (see also the lower panel in Figure 2). 
The line width is significantly broader than those of C1 and C2, but much 
narrower than those of C3 and C4. The line can be identified as 
blended of \ion{Fe}{2} multiplet 191 (1785.26, 1786.74 and 1788.07\AA with 
intensity ratios 20:20:18). After taking into account of different 
instrumental resolution and line blending, the line 
center and width are consistent with the C2 component. 

In order to obtain more physically meaningful fit, we tied the center 
and width of each component to be same for different lines. The flux ratio 
of multiplets is either fixed at its theoretical value or set as a free 
parameter. The [\ion{O}{3}]$\lambda$4959/$\lambda$5007 is fixed at its 
theoretical value of 1/3 for C1, C2 and C3, and is allowed to vary for 
C4. \ion{N}{3}]$1750$ consists of a blend of five lines at 1746.82, 
1748.65, 1749.67, 1752.16 and 
1754.00\AA,  and their relative ratios are fairly constant at density below
 $10^{7.5}$~cm$^{-3}$. We fixed the branch ratios at 0.03, 0.096, 0.502, 0.355, 
0.097 as predicted by CLOUDY (See Appendix 1). \ion{C}{4} has two multiplets with a 
separation of 500 \kms. The  multiplet ratio (1548/1551\AA) is constrained 
in the range 2.0 to 1.0.
We assume that $\lambda$1900 blending features are dominated by \ion{Si}{3}] 
and \ion{C}{3}], rather than by their forbidden counterparts. Justification for 
these assumptions can be found in Appendix 1. To couple with the systematic 
wavelength calibration uncertainty of FOS, which can be up to on diode 
(or ~250 \kms), we allow all UV lines to be systematically 
shifted up to 250 \kms. We assume that each line in the [\ion{N}{2}] and [\ion{S}{2}] 
doublets has the same profile. The [\ion{N}{2}]$\lambda6583/\lambda$6548 ratio 
is set to its theoretical value of 3/1.  The \ion{O}{3}]$\lambda$1660/$\lambda$1666 
doublet ratio is fixed to $1/2.4$ for C3 and C4. The initial values for the 
centroid and width of each component are estimated by fitting individual
lines with a Gaussian: [\ion{S}{2}], [\ion{O}{1}], [\ion{O}{2}], and [\ion{N}{2}] 
for C1; [\ion{Ne}{3}], [\ion{O}{3}] 
and H$\beta$ for C2 and C3; and \ion{C}{4}, \ion{N}{3}], and [\ion{O}{3}] for C4. A total of 
19 emission lines are fitted simultaneously, including \ion{C}{4}, 
\ion{N}{3}], 
Al III, \ion{Si}{3}], \ion{C}{3}], [\ion{O}{2}], [\ion{Ne}{3}], H$\gamma$+[\ion{O}{3}]$\lambda$4363, 
H$\beta$, [\ion{O}{3}]$\lambda\lambda$4959, 5007, [\ion{O}{1}]$\lambda$6302,
H$\alpha$+[\ion{N}{2}]$\lambda\lambda$6548,6584, [\ion{S}{2}]$\lambda\lambda$6717,
6731, \ion{He}{2}$\lambda$1640, \ion{O}{3}]$\lambda\lambda$1660,1666, 
\ion{N}{4}$\lambda$1486 and FeII 191. 
The final fit is shown in Fig. \ref{fig3}.

Balmer decrements are very different for different components from above 
decomposition: C2 has the largest H$\alpha$/H$\beta$ ratio of 12, following 
by C1 (6.2), C3 (5.1) and C4 (4.2). Both of them are considerably steeper 
than 3.1 for an unreddened AGN. If this is interpreted as 
due to dust reddening, then C2 is heavily obscured while the others are 
mildly extincted. In order to see if this is caused by the decomposition 
procedure, we plot H$\alpha$/H$\beta$ at different velocity bin. Here, 
H$\alpha$ profile is obtained by subtracting the best-fit [\ion{N}{2}] model 
from the observed spectrum. Because C1 is well defined and C2 is weak for 
[\ion{N}{2}], [\ion{N}{2}] subtraction should not introduce substantial 
uncertainty. The result is shown in Fig \ref{fig5}, where the data are 
adaptively rebined to ensure the S/N ratio in each bin. Obviously, the 
Balmer decrement varies dramatically across the profile: it rises slowly 
from -4000 to -1000 \kms; then increases sharply from -800 \kms~ and 
reaches its maximum of $H\alpha/H\beta \sim 10$ at the C2 centroid 
and then decreases rapidly to the red wing. Thus large Balmer decrement 
of C2 is independent of the detailed line decomposition. The well defined 
profile of Balmer decrement provides an additional evidence for C2 as 
a distinct component. 

\section{Physical Interpretation of Emission Line Components 
\label{emissiongas}}

We have decomposed empirically the emission lines into four components. 
In this section, we will try to give a physical interpretaion for those 
components based on the physical conditions that are required to 
reproduce the observed emission line ratios. We will use the photoionization 
models as a guideline in the following discussion, but keep in mind that 
shocks may also play certain roles giving the presence of outflows indicated 
by blue-shifted line profiles.  

\subsection{Component 1}

\begin{figure}
\plotone{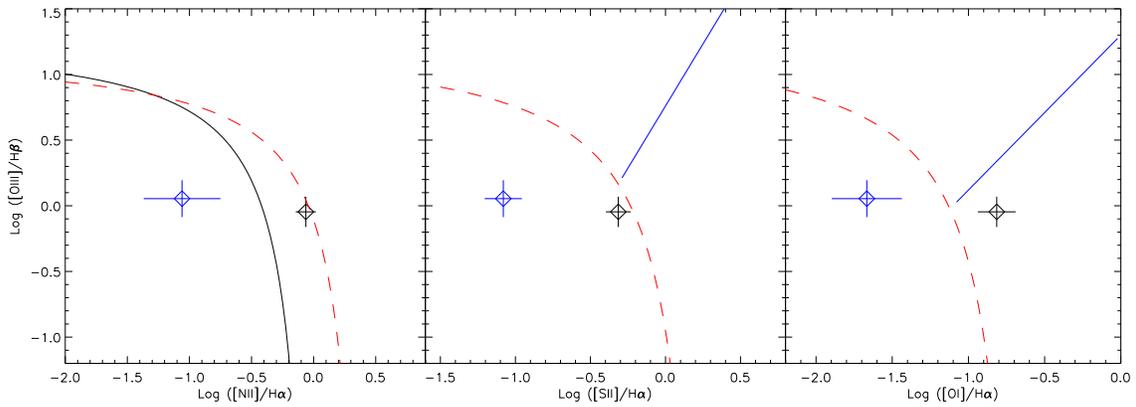}
\caption{BPT diagrams for C1 (black diamonds) and C2 (blue diamonds). 
Black curve represents the empirical boundary separating star-forming 
galaxies from AGNs (Kauffmann et al. 2003), while red curves for the 
theoretical curve of extreme starburst (Kewley et al. 2001). The blue 
straight lines divide LINERs from Seyfert galaxies.}
\label{fig4}
\end{figure}

\begin{figure}
\plotone{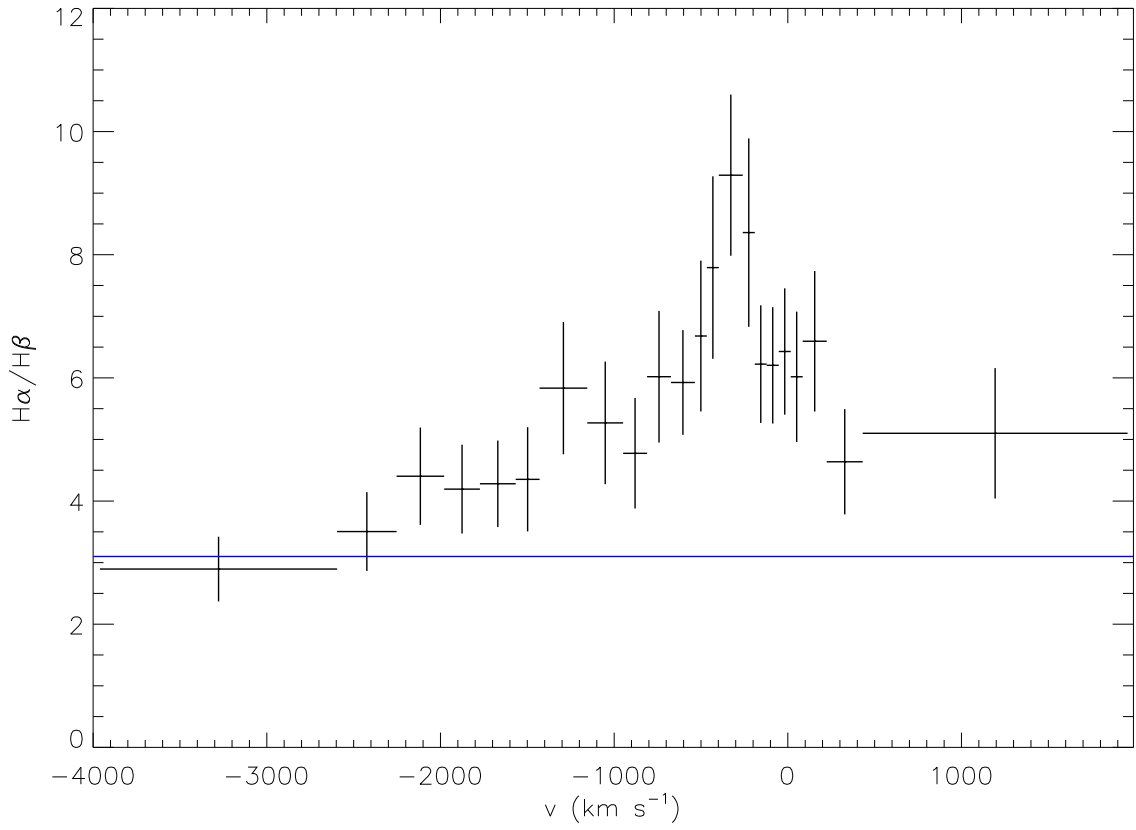}
\caption{Wavelength-dependent H$\alpha$/H$\beta$ ratio over the 
emission line profile. The best-fit [\ion{N}{2}] model has been subtracted 
from the H$\alpha$+[\ion{N}{2}] blend. It shows a clear peak at position of 
C2. The horizon line marks H$\alpha$/H$\beta$=3.1.
}
\label{fig5}
\end{figure}

C1 has doublet ratios of [\ion{S}{2}]$\lambda$6716/[\ion{S}{2}]$\lambda$6731$\simeq$ 1.5 
and [\ion{O}{2}]$\lambda$3729/[\ion{O}{2}]$\lambda$3726 $\simeq$ 0.70, which reach their 
low density limits, and hence the density of the emitting gas should be no larger than 
a few hundred cm$^{-3}$. However, its classification is ambiguous on the conventional 
BPT diagram \citep{bpt81,kewley06}. On the [\ion{N}{2}]/H$\alpha$ versus [\ion{O}{3}]/H$\beta$ 
diagram (Fig \ref{fig4}), C1 is very close to the extreme \ion{H}{2} curve of Kewley et al. (2001). 
It is in the regime for LINER on [\ion{O}{1}]$\lambda$6303/H$\alpha$ vs 
[\ion{O}{3}]/H$\beta$ diagram, but on the \ion{H}{2} side on the [S II]/H$\alpha$ 
vs [\ion{O}{3}]/H$\beta$ diagram.  The Balmer decrement 
$H\alpha/H\beta \approx 6.1\pm 1.6$ indicates substantial reddening 
($E(B-V)=0.61\pm 0.26$~mag, assuming an intrinsic $H\alpha/H\beta=3.0$). 
With such a reddening, the weakness of the corresponding component in the 
UV lines can be explained naturally. With an observed ratio 
[\ion{O}{2}]$\lambda$3727/[\ion{O}{3}]$\lambda5007 \simeq 7.3$, the 
reddening-corrected ratio would be around 14, extreme among the LINERs 
\citep{kewley06}. This puts it on the border between \ion{H}{2} and LINER 
on the [\ion{O}{1}]/H$\alpha$ vs [\ion{O}{2}]/[\ion{O}{3}] diagram. 
This component can be a mixture contribution from \ion{H}{2} and AGN, 
a combination of photoionization and shock-excitation, or gas ionized by 
a diluted AGN continuum (Dopita \& Sutherland 1996).  

\subsection{Component 2}

At first glance, C2 is formed in the star forming regions. [\ion{N}{2}], 
[\ion{S}{2}], and [\ion{O}{1}] emission lines is relatively weak with 
respect to H$\alpha$. Those line ratios fall well in the \ion{H}{2} 
locus on all three BPT diagrams (Fig \ref{fig4}). The gas has a relative 
low metallicity ($Z\sim 0.2-0.3$ Z\sun), according to 
[\ion{N}{2}]/H$\alpha$ ratio \citep{pp04,nagao06}. However, giving the 
high luminosity of the galaxy, the low gas metallicity is not expected. 
In addition, large 
[\ion{Ne}{3}]$\lambda$3869/[\ion{O}{3}]$\lambda$5007$\simeq$1.15 requires 
a rather high gas density ($>10^7$~cm$^{-3}$). Even higher density is required 
if the large Balmer decrement of C2 is attributed to dust reddening. In this 
case, weakness in [\ion{N}{2}], [\ion{S}{2}] and [\ion{O}{1}] is naturally 
explained because they are collisionally de-excited, and BPT diagrams no long 
provide diagnostics for the ionizing source, as such the ionizing source can 
be AGN as well. 

The strong \ion{Fe}{2} 191 emission in component 2 is a puzzle. First, its 
emission region cannot be reddened as severely as indicated by the apparent 
Balmer decrement, which suggests an intrinsic dust reddening of E(B-V)=1.2 
mag using Calzetti's extinction curve for an intrinsic H$\alpha$/H$\beta$=3.0. 
This indicates either a large intrinsic H$\alpha$/H$\beta$ or a separated 
\ion{Fe}{2} 191 emission region. Second, because of lack a similar 
component in other UV lines, \ion{Fe}{2} 191 must be produced in a region 
shielded from intensive Uv ionizing radiation. While whether this can be 
consistent with the prominent emission in [\ion{O}{3}] and [\ion{Ne}{3}] 
needs to be further studied, a large column density may explain this. Third, \ion{Fe}{2} 
191 is strong while optical \ion{Fe}{2} emission is weak. The dominance of 
emission from high excitation levels indicates that either it is formed in a 
high temperature and dense HI region or/and through UV photon excitation 
by the continuum radiation (PCR). A high temperature and dense HI region 
may help to explain H$\alpha$/H$\beta$ ratio in this component through 
collisional enhancement of H$\alpha$ emission.   
Because this component is blueshifted by ~350 
km~s$^{-1}$ with respect to the systematic velocity, interaction between 
the outflow and interstellar medium may be a viable heating source. 
PCR process had been suggested for the \ion{Fe}{2} emission from the wind 
of spectroscopic binary KQ Puppis (Redfors \& Johansson 2000). In order 
to make photon-pumping effective, the gas must see a strong soft UV 
photons and has large internal velocity gradient. We speculate that an 
ionizing continuum filtered by a thin partial ionized source may mimic 
the soft UV photons (see a similar idea in Collins et al. 2009).  The 
PCR process would also produce \ion{Fe}{2} 193 emission lines at 1459, 
1465, 1474\AA~ but the spectrum quality does not allow us to  assess 
their existence.

Therefore, we tentatively interpret this component as from a dense region 
: The large Balmer decrement is 
interpreted as due to radiation transfer effect or caused by collisional 
excitation in the warm dense HI region, rather than dust reddening; 
\ion{Fe}{2} 191 is produced either in a high temperature warm HI region 
or through PCR process; [\ion{N}{2}], [\ion{S}{2}] are suppressed by 
collisional de-excitation. The blueshift of C2 (350 \kms) relative to 
the systematic velocity can be readily interpreted as from an 
outflow with the far-side, redshifted counter-flow being obscured. 
However, we cannot rule out the possibility that C2 consists of emission 
lines from a dusty star forming region and an additional dense warm partially 
ionized gas component.
 
\subsection{Component 3}

C3 has strong Si III], C III], Al III, [\ion{O}{3}], [Ne III] and  Balmer lines, 
but is undetected or very weak in \ion{C}{4}, \ion{N}{3}], [O I], 
[O II], [S II] and [N II]. Relatively high ionization and excitation suggest 
that this component is produced by AGN photoionization. \ion{O}{3}]1665/[\ion{O}{3}]5007, 
\ion{C}{3}]1909/[\ion{O}{3}]5007 and [\ion{O}{3}]5007/H$\beta$ suggest a density of 
$10^{7.5-8}$~cm$^{-3}$, while [\ion{O}{3}]$\lambda$4363/$\lambda$5007 and 
[\ion{Ne}{3}]3896/[\ion{O}{3}]5007 indicate a somewhat lower gas density about 
$10^7$~cm$^{-3}$. As [\ion{O}{3}]4363 is badly blending with H$\gamma$ line and 
the flux of [Ne III] is sensitive to the way of modeling the red wing, the 
latter two should be used with care. Therefore, we believe that 
$10^8$ cm$^{-3}$ is more likely the real value. 

However, the large line ratio of \ion{Si}{3}]/\ion{C}{3}]$\sim$1 can 
only be reproduced with a high density $n_H>10^{9.5}$ cm$^{-3}$, if the gas has 
abundances being the solar or the scaled solar values (See Appendix 2).
This may be explained in terms of at least two emission line regions 
with different densities. Such models, however, have certain drawbacks as the 
following. If the high and low density gases are at the same distances
and see the same continuum, the large difference in density would result in
an ionization parameter for the low density gas being three orders of magnitude 
higher than that for the high  density gas; this would lead to strong coronal 
emission lines, such as [Fe VII], which are not seen in the observed spectrum. 

A single zone model can be constructed if Si element is enhanced relative to 
C. Overabundance of $\alpha$-elements relative to C is predicted by the 
starburst chemical evolution models for quasar host galaxies (Haimann \& 
Ferland 1993). Considering the fact that Q1321+058 experienced a recent 
starburst (see next section), this chemical enrichment route is plausible. 
In their models, more massive stars are produced with respect to 
star formation in the Galaxy, thus leads to both fast metal enrichment 
and a high final metal abundance ($\geq$ 10 Z\sun). $\alpha$-elements are 
enhanced relative to C because C-enrichment from intermediate mass stars 
has a finite time delay with respect to 
SN II+Ib eruption, which produces most $\alpha$-elements. It should be 
noted that the relative abundances of different metals depend on the 
metallicity. With increasing metallicity, N relative to O increases 
very fast at all abundances, while Fe and $\alpha$-elements rise 
substantially. Lack of \ion{N}{3}], \ion{N}{4}] and \ion{C}{4}, we can only 
put an upper limit on the gas metallicity to 3 solar value based on 
\ion{N}{3}]/\ion{C}{3}] because other lines are not sensitive to the 
metallicity. 

Over the density range considered in Appendix 2, \ion{Si}{3}]/\ion{C}{3}] 
provides a good 
diagnostics of the ionization parameter. The observed \ion{Si}{3}]/\ion{C}{3}] 
ratio can be reproduced with two branches of ionization parameters: an upper 
branch with $\log U \sim -0.5$ and a lower branch $-2.0 \le \log U < -1.5$,  
at a column density $\log N_H ({\mathrm cm}^{-2})=21$.  At larger column 
densities, we find that only the low ionization branch is possible, however. 
In the density range, \ion{C}{3}]/\ion{C}{4} is sensitive both 
to the ionization parameter and to the column 
density. The lower limit on \ion{C}{3}]/\ion{C}{4} suggests a large column density 
$\log N_H\geq 22$ and a low ionization parameter $\log U<-2.2$, thus the 
lower branch solution is preferred. With $n_H \sim10^{7.5}$~cm$^{-3}$, 
$\log U\sim -2.0$ and $\log N_H\geq 22$, the observed line ratios can be 
reproduced roughly. The relatively high density can also explain 
naturally the weakness of this component in [\ion{O}{2}], [\ion{N}{2}], 
and [\ion{S}{2}]. 

With a relative large emission line width and close to systematic velocity, 
this component may be gravitational bound optical thick clouds ionized by 
the AGN continuum. We speculate that it is the intermediate emission line 
region between BLR and NLR.  

\subsection{Component 4}

C4 is present in all permitted and semi-forbidden lines, as well as
the high-ionization forbidden lines [\ion{O}{3}] and [\ion{Ne}{3}]. 
The line ratios [\ion{O}{3}]4363/[\ion{O}{3}]5007, 
\ion{O}{3}]1665/[\ion{O}{3}]5007, 
\ion{C}{3}]1909/[\ion{O}{3}]5007, and 
H$\beta$/[\ion{O}{3}]5007 give consistent densities around $10^7$cm$^{-3}$, 
while [\ion{Ne}{3}]3869/[\ion{O}{3}]5007 suggests a somewhat lower density ($10^{6.6}$
cm$^{-3}$). These line ratios are only weakly dependent on the metal 
abundances or gas column density, thus we believe the density is quite 
robust. As for C3, at this density range, a large \ion{Si}{3}]/\ion{C}{3}] 
requires that Si is over-abundant relative to C, and this may be 
interpreted as the starburst abundances of Haimann \& Ferland (1997). 
Large \ion{N}{3}]/\ion{C}{3}] ratio can be reproduced only with a very supersolar 
abundance ($Z\sim10 Z_\odot$; see Fig \ref{fig7}) because nitrogen abundance 
increases with the average metallicity ($Z$) as $\propto Z^2$ at high 
$Z$ in their models. 

As seen in Fig 7, we are unable to reproduce quantitatively all line 
ratios with a single constant density models described in Appendix B, 
but we believe this can be solved by adjusting the input ionizing 
continuum. Metallicity independent line ratios, \ion{C}{3}]/\ion{C}{4} 
and \ion{N}{3}]1750/\ion{N}{4}]1846 suggests different ionization parameters 
$\log U\simeq -1.5$, and $\log U\simeq -2.5$ for a range of gas column 
densities. Since the ionization potential of $N^{2+}$ ($N^{+}$) is only 
slightly larger than $C^{2+}$ ($C^{+}$), any smooth change in the continuum 
slope in this regime is not likely to cause such big difference. However, we 
note that the ionization potential (55.45eV) of $N^{2+}$ is just above that of 
\ion{He}{2} ionization potential, while that (47.89eV) of $C^{2+}$ is below 
the \ion{He}{2} ionization potential. If the continuum incident on the gas has 
already been filtered by a relative highly ionized gas that is 
optically thick to helium ionizing photons, such ionizing continuum 
will have less N$^{2+}$ ionizing photons relative to  C$^{2+}$ ionizing 
photons, resulting more less $N^{3+}$ ions. It would be interesting to 
see if such a model can reproduce \ion{N}{3}]/\ion{N}{4}] and 
\ion{C}{3}]/\ion{C}{4} ratio at the same time, but throughout 
photo-ionization calculations are beyond the scope of this paper. 
We only want to point out here that once such an ionizing continuum 
is adopted, the ionization parameter will be close to, but  
substantially larger than, the higher ionization paramter estimate 
based on \ion{C}{3}]/\ion{C}{4} ratio. 
Therefore, we interprete C4 as a dense outflow drivening by AGN.

\section{Discussion}

\subsection{Outflows}

In Q 1321+058, we have detected emission lines from two outflow 
components, C2 and C4. The broad component C4 with a radial velocity 
1650\,\kms\, has been interpreted as dense, metal enriched outflow 
photoionized by AGN continuum, probably filtered by an ionized 
absorber in last section. The narrow component C2 is also produced 
by dense gas, but its nature is less clear. Therefore, we will focus 
on the mass outflow rate and kinematic power of the high velocity outflow
 C4 in this section first, and then discuss 
briefly the C2 component.
    
We first estimate the mass in the outflow for C4. The \ion{H}{2} mass of the outflow 
can be estimated from the H$\alpha$ luminosity as follows
\begin{equation}
M=\frac{1.4 m_H L_{H\alpha}}{n_e \alpha^{eff}_{H\alpha} h\nu_{H\alpha}}
 = 450\times \left(\frac{L_{H\alpha}}{10^{42} {\rm erg~s}^{-1}}\right)\left(\frac{n_e}{10^7 {\rm cm}^{-3}}\right)^{-1} M_{\rm \odot}
\label{massestimate}
\end{equation}
assuming case-B recombination with a temperature of 10$^4$ K (Table 4.2 
of Osterbrock 1989). The factor 1.4 accounts for Helium mass. With the 
H$\alpha$ luminosity of 1.5$\times 10^{42}$ \ergs, we obtain 
$\sim$680$(n_e/10^7)^{-1}$~M\sun~ for the high velocity outflow. 
This mass should be regarded as lower limits to the total mass in 
consideration of the existence of possible HI regions. We 
can estimate the mass of emission line gas using the best model 
in section \ref{emissiongas} as well. The model gives a similar value 
$\sim 400$ M\sun.  

For the single zone photoionization models discussed in \S3.1, we estimate 
the distance of the outflow to the AGN,
\begin{equation}
R=\sqrt{\frac{Q_{ion}}{4\pi n_HUc}}=150 \sqrt{\frac{10^{7}}{n_H}\frac{0.01}{U}} {\mathrm pc}. \label{dist}
\end{equation}
Assuming a typical quasar SED (Richards et al. 2006), the total ionizing 
flux of Q 1321+058 is estimated to be 8$\times$10$^{57}$~photons~s$^{-1}$ 
(see \S \ref{activity}). Using the density and ionization parameter 
estimated in \S3.1, we find distances of $\sim 80~pc$ for C4, which is one 
order of magnitude smaller than the size of NLR in quasars 
of similar bolometric luminosity (Benert et al. 2005). Note the distance 
is not very sensitive to the ionization parameter used.  

The mass loss rate is small compared to the mass accretion rate onto 
the supermassive black hole. The emission line gas is visible for a 
dynamic time scale of $\Delta R/v_r$, where $\Delta R$ is the visible 
radial length of the outflow, and $v_r$ is the observed radial velocity. 
By combining with the mass in Eq \ref{massestimate}, we obtain 
an estimate of the mass loss rate,
\begin{equation}
\dot{M}=\frac{M v_r}{\Delta R}=2\times 10^{-2}
\left(\frac{L_{H\alpha}}{10^{42}~{\rm erg~s}^{-1}}\right)
\left(\frac{10^7~{\rm cm}^{-3}}{n_e}\right)
\left(\frac{v_r}{2000~{\rm km~s}^{-1}}\right)\left(\frac{100~{\rm
pc}} {\Delta R}\right) M_{\rm \odot}~{\rm yr}^{-1} \label{rate}
\end{equation}
Note with the column density estimated in the section \S 
\ref{emissiongas}, the corrections due to HI gas are within a 
factor of two. The mass loss rate is order 
of 0.03 M\sun~yr$^{-1}$ if $\Delta R/R\sim 1$. Therefore, the mass 
loss rate for these outflow is not important with respect to the 
mass accretion rate onto the supermassive black hole (see next section). 

The C4 outflow does not provide sufficient kinetic power to 
account for the feedback that is supposed to heat the cold gas in 
the host galaxy up to the escape velocity. The kinetic power 
with the line emitting gas is only on the order of $\sim 10^{-6} 
(100 pc/\Delta R)$~M\sun 
$c^2$~yr$^{-1}$, or 10$^{-5} (100 pc/\Delta R) $ of the AGN bolometric 
luminosity according to the mass loss rate in Eq \ref{rate} and the observed 
outflow velocity, while a few percent is required for the flow to be 
energetically important (e.g., Di Matteo et al. 2005). A similar 
conclusion has been reached by modelling absorption lines for nearby 
Seyfert galaxies (Crenshaw, Kraemer \& George 2003).

However, instead to conclude that outflows are energetically un-important 
for AGN-galaxy feedback, we argue below that there is an indication for much 
larger kinetic power than the above estimate. 
As we have seen in the last section that the column density of the emission gas 
is fairly low, of order of $10^{21}$~cm$^{-2}$. A gas cloud with such column 
density, density and typical temperature of photoionized gas is not bounded 
by self-gravity, thus it will either disperse away on the time scales of only 
10 years  (i.e., $\Delta R\sim 2\times 10^{-2}$ pc)
or be confined by external pressure. In the former case, continuous formation 
of the line emitting gas is required, and the mass outflow rate would be 40 
M\sun~yr$^{-1}$, and total kinetic power would be three magnitude higher than 
the above estimate, which is sufficient to explain AGN-galaxy feedback. In the 
latter case, the confined medium can be hot phase gas. Since the filling 
factor of the warm gas is extremely small 
($\sim 10^{-10}\times\frac{100~{\rm pc}}{\Delta R}$), 
the bulk of the gas will be in the hot phase, and both the mass loss rate 
and the kinetic power may be much larger than above estimate. 
Hot gas has been detected in nearby 
ultra-luminous infrared AGN including NGC 6240 (Komossa et 
al. 2003). Future high spatial resolution X-ray observations can be 
used to examine if such a hot X-ray component presents in this 
quasar as well. 

The covering factor of the emission line regions can be estimated from 
the ratio of H$\alpha$ to the ionizing photon flux assuming Case-B 
recombination if the emission line region is optically thick to the 
ionizing photons. It turns out that the covering factor of C4 
is as small as 6$\times 10^{-5}$. Similar numbers 
(order of $\sim 10^{-4}$) are also obtained by comparing the \ion{C}{3}]1909, 
[\ion{O}{3}]$\lambda$5007 or H$\alpha$ luminosity with the best photoionization 
model in the last section in combination with above distances. For 
comparison, the covering factor of the NLR in quasars is typically a few 
percent according to simple photon-counting estimation (e.g., Netzer 
\&~Laor 1993). Thus, the AGN outflows in Q 1321+058 has a covering 
factor 100 times smaller than a typical NLR. 
 
In passing, with a maximum velocity extending to $\sim$4000 \kms~ 
and emission lines of a wide ionization range, Q1321+058 would appear 
as a LoBAL quasar if the line of sight to the nuclear continuum is 
intercepted by the outflow. We speculate that Q 1321+058 is an 
obscured, high luminosity version of MKN 231, an extensively studied 
infrared luminous low ionization BAL QSO with an extended outflow 
region on kpc scale \citep[e.g.][]{lipari05}. The requirement of a 
soft ionizing continuum for C4 is also consistent with weak 
X-ray emission from BAL QSOs.

We would like to point out that the properties of outflow in C2 component 
is very uncertain. First, \ion{Fe}{2} emission requires an extensive 
warm HI region, but its column density cannot be properly constrained 
without extensive modelling of the FeII emission. Thus, Eq (\ref{rate}) 
will give only a lower limit to the mass loss rate. Second, if a large 
fraction of H$\alpha$ photons are produced through collision excitation 
as is required to explain its large intrinsic Balmer decrement, Eq 
(\ref{rate}), on the other side, Eq (\ref{rate}) is no long valid. 
Finally, it has yet to be demonstrated that collisional excitation 
can fully explain the large H$\alpha$/H$\beta$. Otherwise, dust 
extinction may be important. 

\subsection{Quasar Activity and Star Formation History \label{activity}}

\begin{figure}
\plotone{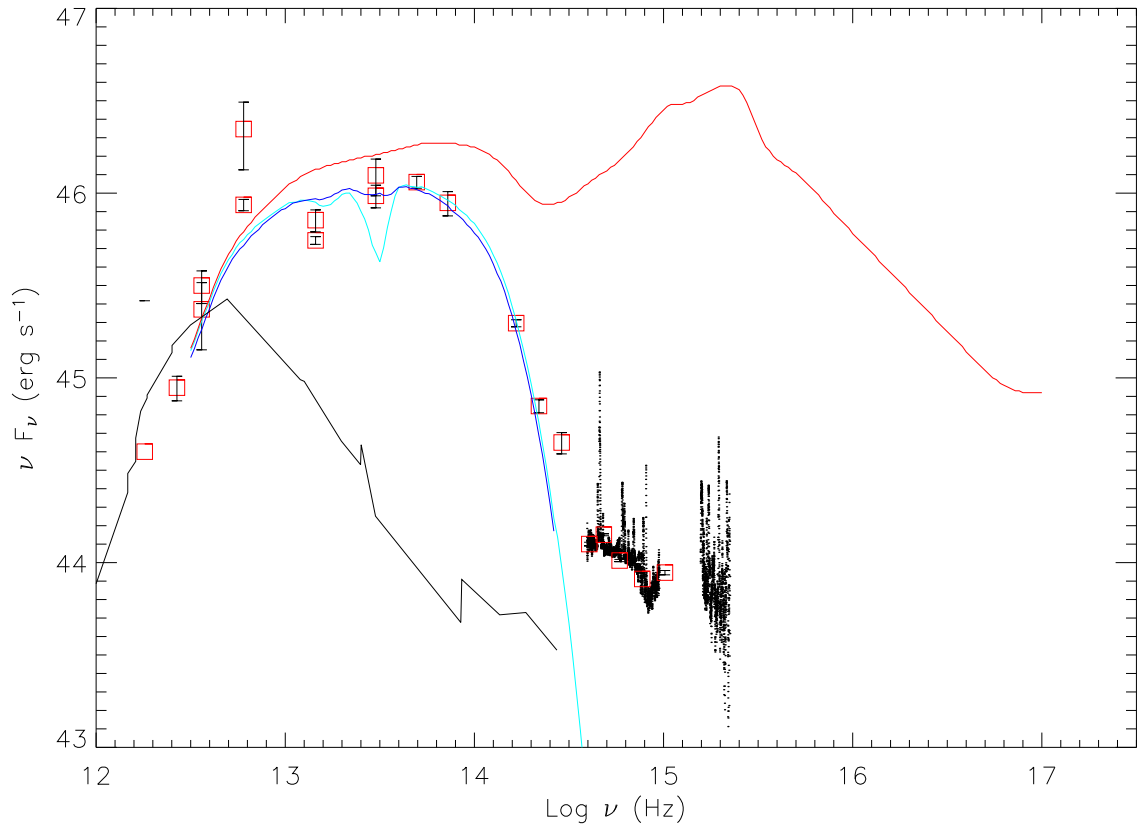}
\caption{Spectral energy distribution from far-infrared to ultraviolet
for Q1321+058. The infrared SED of NGC 6240 is shown in solid line, the
template of infrared-luminous quasars in Richards et al. (2006) 
in red, while the quasar template reddened by E(B-V)=4.5 in blue.
}
\label{fig6}
\end{figure}

The broad band spectral energy distribution (SED) of Q 1321+058 is 
plotted in Fig. \ref{fig6}. The near-infrared data are taken from 
the two micron all sky survey
\citep[2MASS][]{skr06}. The mid- and far-infrared data are
drawn from observations by the infrared space observatory (ISO)
\citep{kur03} and the IRAS faint source catalog \citep{mos90},
respectively. Q1321+058 was also observed with Spitzer IRAC and 
IRS instruments (Farrah et al. 2007; Sirocky et al. 2008). 
The SDSS and HST FOS spectra, as well as the SDSS photometric data 
corrected for Galactic extinction are also plotted.

The mid-infrared bump is very prominent in the SED, that rises 
steeply from near- to mid-infrared, and then flattens and declines 
in far-infrared. There is no apparent dip or peak around 9.7 micron 
in the source rest frame. Lack of strong emission or absorption 
feature is confirmed by Spitzer IRS observation, which reveals only 
weak silicate absorption feature\citep{sirocky08}. The flat middle 
to far-infrared spectrum and a lack of prominent PAH features 
suggest that the bulk of the mid-infrared emission is powered by 
the hidden quasar \citep{weedman05,hao07}. To compare the observed 
SED with that of red quasars, we redden the average SED of infrared 
luminous quasar from Richards et al. (2006) through a uniform dust 
screen with $E(B-V)$=4.5 using the extinction curve of Weingartner 
\& Draine (2001; 30ppm C in PAH), which can reproduce
approximately the extinction curve to the Galactic center. The
result is also shown in Fig. \ref{fig3}. This over-simplified model 
can reproduce the overall shape of the infrared SED except for 
predicting a strong dip around 10 $\mu$m due to small grains 
\citep[see also][]{ld93}, which can be eliminated with a detail 
treatment of radiative transfer \citep[e.g.][]{siebenmorgen04}.

The 12$\mu$m  to total observed [\ion{O}{3}] flux ratio is 50 time 
higher than the average Seyfert galaxies (Haas et al. 2007). Even 
after correcting for reddening, this ratio is still more than a factor of 
ten higher. However, we consider that this can be attributed to 
weak [\ion{O}{3}] emission rather than the dominance of mid-infrared 
flux from the star-formation region for the aforementioned reason. 
Note weak [\ion{O}{3}] emission is a common characteristics of low ionization 
BAL QSOs (Boroson \& Meyers 1993). 

The total luminosity in the infrared is $1.9\times10^{46}$ ergs~s$^{-1}$, 
which makes Q 1321+058 an ULIR quasar. If Q 1321+058 bears an intrinsic 
SED of a typical quasar, the bolometric luminosity is estimated to be 
$3.9\times10^{46}$ ergs~s$^{-1}$, using the quasar template fit to 
infrared SED and integrating in the optical to UV portion
\footnote{Infrared emission is the dust reprocessed light from the optical 
and ultraviolet continuum, thus should not be counted in the bolometric 
luminosity calculation for an unobscured quasar.}. However, if the 
covering factor of 
dust in this object is larger than the average value of quasars, the bolometric 
luminosity is then over-estimated from the above fit. A 
conservative estimate can be given by assuming that all of the UV 
to X-ray continuum emission is absorbed and re-emitted isotropically 
in the infrared. This yields a bolometric luminoisty a factor of two smaller. 
The mass accretion rate is thus $\sim$ 3-6 M\sun~yr$^{-1}$, for the typical 
radiative efficiency of 0.1. If the quasar is accreting at close to the 	
Eddington limit, the black hole mass is estimated to be around 10$^8$ M\sun.

The SFR for this object may be estimated in several ways. We have
found that UV spectrum can be fitted by a young stellar population. 
With the age and mass of the young stellar population, we estimate 
a SFR of about 450-230 M\sun~yr$^{-1}$. However, as we noted this 
number is only valid to the order of magnitude giving our 
oversimplified assumption about the model. Alternatively, using the UV 
continuum luminosity, the SFR calibrator of Madau et al. (1998) and 
assuming an extinction E(B-V)=0.26, we obtain a SFR of only 
45~M\sun~yr$^{-1}$. The discrepancy arises because the SSP model contains 
1 Myr population only, while older stellar populations still 
contribute to the observed UV flux. The SFR can also be estimated from 
the far-infrared luminosity and radio power. Assuming most of the far-infrared 
luminosity is powered by the starburst\citep[e.g.,][] {schw06}, we find an 
upper limit to SFR of about 270 M\sun~ yr$^{-1}$, similar to that seen in 
NGC 6240 (e.g., Pasquali, Gallagher \& de Grijs 2004). Q 1321+058 was 
detected in both NVSS and FIRST with a flux density of 
$S_{1.4GHz,NVSS}=4.9\pm 0.5$ mJy and $S_{1.4GHz, FIRST}=4.5\pm 0.3$ mJy. 
Assuming that the radio emission ($6\times10^{30}~erg~s^{-1}~Hz^{-1}$) all 
comes from the \ion{H}{2} region, we obtain an upper limit of $SFR \lesssim 330~
M_{\odot}~yr^{-1}$. These values are consistent with that derived from the UV 
spectral fit. The SFR can be also estimated with the PAH luminosity 
\citep[e.g.]{wu05}. \citet{cao08} detected a 6.2$\mu$m PAH luminosity 
of 1.2$\times10^{42}$~erg~s$^{-1}$. Following Hernan-Caballero et al. 
(2009), we obtain a SFR of only 4.6 M\sun~yr$^{-1}$. Finally, the SFR can be 
estimated from  H$\alpha$ luminosity. An upper limit can be derived if 
C1 is assumed to be ionized by young stars. The extinction corrected 
H$\alpha$ luminosity for C1 are 1.8$\times10^{42}$ erg~s$^{-1}$. 
According to the calibration of Madau et al. (1998), the SFR is 
14 M\sun~yr$^{-1}$. SFR will be one order of magnitude larger if C2 is 
from star-formation region. 
The very different SFR derived from different methods 
demonstrate the difficulty of estimating SFR in the presence of AGN activity.

With all above estimates, SFR does not likely exceed hundreds M\sun~yr$^{-1}$ 
, while we have found a mass accretion rate of 3-6 M\sun~yr$^{-1}$ to the black hole according to its bolometric 
luminosity. The $\dot{M}/SFR$ is more than an order of magnitude larger 
than that required for the strict co-evolution of the black hole and 
bulge. However, the black hole to stellar mass ratio can still be 
consistent with that found in local spheroidal galaxies 
(Gebhardt et al. 2000). The mass of the intermediate-aged stellar 
population in Q 1321+058 is $\sim 6\times 10^{10}$ M\sun~ from our 
analysis of the SDSS and HST 
spectra. If the bolometric luminosity does not exceed the Eddington 
luminosity, a lower limit to the black hole can be set to $10^8$ M\sun. 
It was argued that ultraluminous infrared quasars are accreted at 
the Eddington limit rather than fuel limited (Hao et al. 2005). 
Then a lower-limit on a mass ratio of the black hole to the intermediate 
and young stellar populations to be $\sim 2\times 10^{-3}$, in coincidence 
with that for local spheroidal galaxies (Gebhardt et al. 2000). 

\section{Summary}

We have performed a detailed analysis of the optical--UV 
spectrum and the broad band spectral energy distribution 
of Q 1321+058. Our analysis confirms that it is an obscured quasar 
with its bulk energy output in mid-infrared. 
Its optical--UV  spectra show complex emission lines. 
We identified four components: a narrow component at the systematic 
velocity (C1), a narrow component at velocity -380 \kms~(C2), and 
two broad components at -80\kms~(C3) and 1650\kms~(C4), respectively. 
C1 shows a LINER/composite spectrum, which is common among ULIRGs. Both 
C2 and C4 can be interpreted as dense outflows with the back-sides are 
obscured. We speculate that C3 comes from an intermediate line region 
between BLR and NLR.    

A comparison of the measured emission line ratios with photo-ionization 
models suggests that C4 outflow has a gas density $n_H\sim 10^{7}$~cm$^{-3}$, 
column density $N_H\sim 10^{21}$~cm$^{-2}$, an $\alpha$-enriched  
super-solar metallicity of $Z\sim 10~Z$\sun~for starburst galaxies. It 
is located at a distance about a hundred parsecs from the central 
continuum source. The velocity range, ionization level, and column 
density derived from the emission lines suggest that Q 1321+058 might 
be a low ionization broad absorption line (LoBAL) quasar viewed in an 
``unfavored'' direction, with the outflows being part of the 
otherwise LoBAL region. The apparent total mass loss rate in 
the C4 outflow is small, thus the outflow associated with line emitting 
gas does not have sufficient energy to remove the ISM of host galaxies,
quenching both star-formation and accretion process.
But we argued that the optical emission line region may trace 
only a small amount of quasar outflows, thus the actual mass loss rate 
and kinetic power may be much larger. The covering factor of the outflow
is very small. 

The optical and UV continuum can be well modeled with a young ($\sim
1~Myr$) plus an intermediate age ($0.5-1$ Gyr) stellar population. 
The latter population has a mass around a few $10^{10}$ 
M\sun, suggesting for a recent building of a massive galaxy 
following the merger of two gas rich galaxies. Fast 
stellar mass building is also consistent with the metallicity that 
required to explain the C3 and C4 line ratios. We obtain very different 
SFR or its upper limits, independently from UV continuum, radio, 
far-infrared and emission line luminosity in the range of a few to 
serval hundred M\sun~yr$^{-1}$. We estimate a bolometric luminosity 
of (2-4) 10$^{46}$~\ergs~for the quasar, or a mass accretion rate of 
3-6 M\sun~yr$^{-1}$ for a typical 
efficiency of 0.1. The SFR to mass accretion ratio is more than one order 
of magnitude lower than that is required for the co-evolution of black 
hole and spheroid. If the black hole is accreted at the Eddington 
rate, the black mass to the stellar mass ratio will be coincident with 
that defined by local spheroid galaxies.

\acknowledgements 
We thank the anonymous referee for constructive comments which lead 
to significant improvement in the presentation. This work is supported 
by Chinese Natural Science Foundation through CNSF-10233030 and 
CNSF-10573015, and by the Knowledge Innovation Program of the Chinese 
Academy of Sciences, Grant No. KJCX2-YW-T05. H. Y. Zhou acknowledges 
the Chinese NSF support through NSFC-10473013, and the support from 
NSF AST-0451407 and AST-0451408, NASA NNG05G321G and NNG05GR41G, and 
the University of Florida.

\appendix
\section{Notes on the fit to [\ion{O}{3}], \ion{N}{3}] and \ion{Si}{3}] emission lines}

Initially, we fit the four components to the 
[\ion{O}{3}]$\lambda\lambda$4959,5007 blend, and for each component the
doublet ratio is fixed to 1/3. The fit is not satisfactory. In
particular, there is an excess in the blue component in
[\ion{O}{3}]$\lambda$4959. This indicates that the doublet ratio of one or
more components is not the theoretical value, or there is
contribution from other emission line(s). Potential contaminating
lines in the regime can be [\ion{Fe}{7}]$\lambda$4989, \ion{He}{1}$\lambda$5016,
[\ion{Fe}{2}I]$\lambda\lambda$4988,4931 or the \ion{Fe}{2} blend. Since the \ion{Fe}{2} bumps
around 4700 \AA~and 5100 \AA~is not visible, we will not consider
this as a likely possibility\footnote{\ion{Fe}{2} forbidden transitions in
this band is weak \citep{veron04}}. Our grid photo-ionization
calculation suggests that the \ion{He}{1} emission should be no more than 
2\% of H$\beta$ in strength 
(for a wide range of parameters as discussed  in Appendix 2),
and is too weak to account for the excess. [\ion{Fe}{7}] $\lambda$4989
can be fairly strong for a high ionization parameter and a low column
density in photoionization models, but [\ion{Fe}{7}]$\lambda$6087, which
is not detected in our spectrum, should be a factor of at least 2.7
stronger for all models. 
Therefore, it is likely that the excess is
due to the [\ion{O}{3}]$\lambda$4959 emission. To incorporate this, we leave
the [\ion{O}{3}]$\lambda$4959/$\lambda$5007 ratio of the blue component as
a free parameter. This gives fairly good fit with the ratio of 0.46.
Therefore, in the subsequent fit, the [\ion{O}{3}] doublet ratio is
allowed to vary freely for C4 and is fixed at 1/3 for other
components.

The \ion{N}{3}]$\lambda$1750 emission is a blend of five lines, and 
the multiplets ratios depend on both gas temperature and density 
\citep{dwi95}. Using the grid photo-ionization models described in 
Appendix 2, we estimate that this feature is dominated by 
\ion{N}{3}]$\lambda$1751 and \ion{N}{3}]$\lambda$1752, and other 
lines of the multiplets account for less than 10\% only. The line 
ratios are only weakly depends on the ionization parameters, we 
fixed the multiple ratios at the model value in the vicinity of 
the best model density. 

Both \ion{Si}{3}]$\lambda$1889 and [\ion{Si}{3}]$\lambda$1883 may 
contribute to the $\lambda$1888 blend. At a low density, [\ion{Si}{3}] 
is the main contributor
to this feature, while at a  density above 10$^6$~cm$^{-3}$ this
feature is dominated by \ion{Si}{3}]. The situation is similar for the
$\lambda$1909 blend of [\ion{C}{3}]$\lambda$1907 and \ion{C}{3}]$\lambda$1909.
As aforementioned, the \ion{Si}{3}/\ion{C}{3} ratio requires this
feature being dominated by semi-forbidden transitions. Therefore, 
we take wavelengths 1889\AA~and 1909\AA~for \ion{Si}{3} and \ion{C}{3}, respectively, 
in the following analysis.

\section{Photoionization Models}

\begin{figure}
\plotone{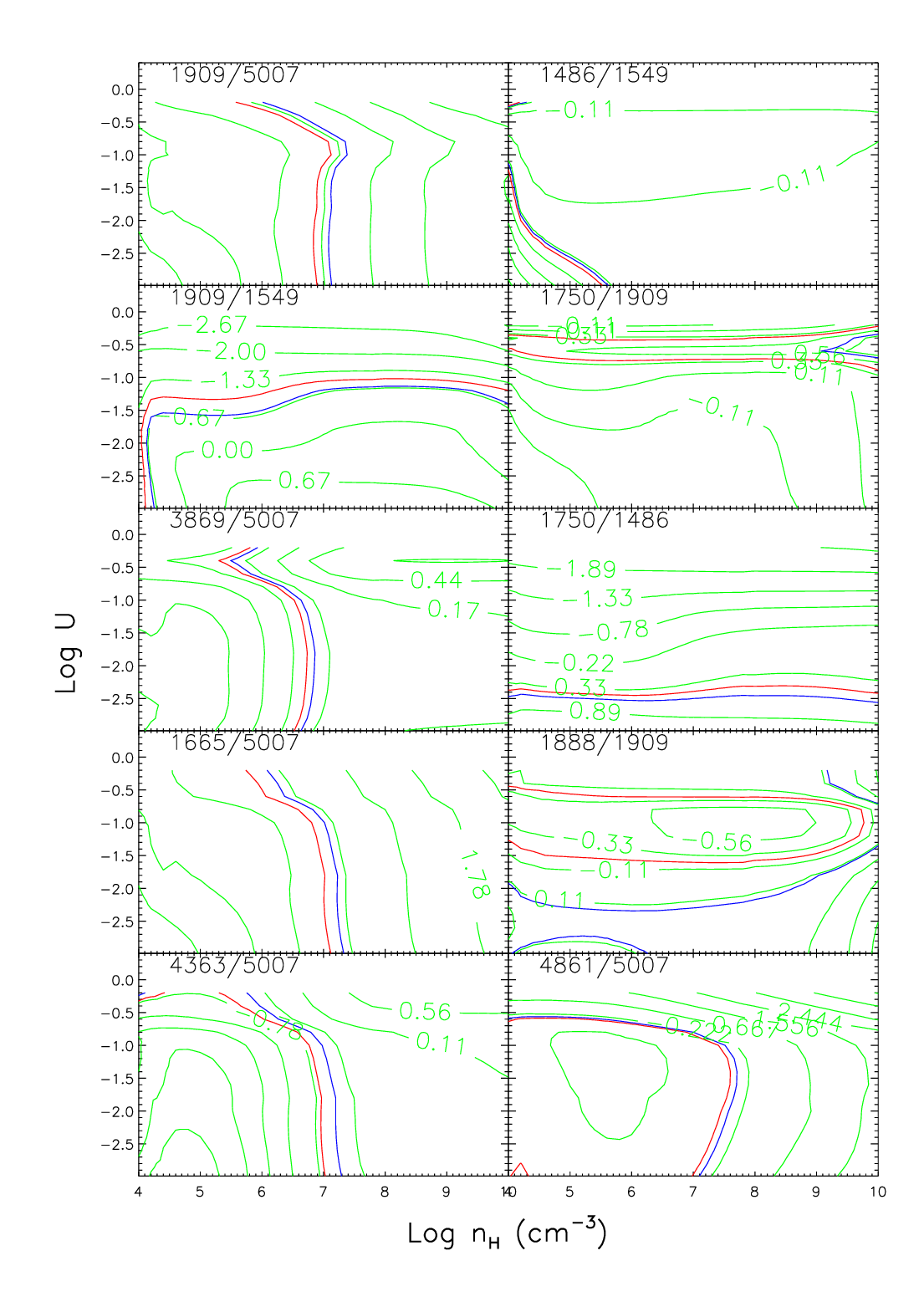}
\caption{Contours of emission line ratios in the ionization parameter 
-- density space calculated using the Cloudy C06.02 (Ferland 1998) for 
metal abundance Z=10Z\sun and a column density $N_H=10^{21}$ cm$^{-2}$. 
The regimes for C4 is between two colored lines (green and blue) after 
taking account for the one sigma error bar for the observed line ratios. 
\label{fig7}}
\end{figure}

We have computed large grid constant-density models using the
photoionization code CLOUDY (version 06.02, last described by
Ferland et al. 1998). The range of gas density considered is $10^{2-10}
~cm^{-3}$, and the range of ionization parameter considered is
 $-3.5<\log U < 0.0$. We consider five column densities of $\log
N_H$[cm$^{-2}$]=21, 21.5, 22.0, 22.5, and 23.0, and four metal
abundance values: 1, 3, 5 and 10 $Z$\sun. Two metal enrichment 
schemes are considered: all metal abundances are scaled-up solar 
values, and metal enrichment in a starburst (Haman \& Ferland 1997). 
We choose a  typical AGN ionizing continuum (Korista et al. 1997). 
We examine the line ratio contours on the dual-parameter planes 
for the plausible regimes (an example of such contours is shown 
in Fig \ref{fig7}). We find that: (1) with a solar and 
scaled solar metallicity, large \ion{Si}{3}]/\ion{C}{3}]$\sim 1$ can be 
reproduced only at a high density $n_H>10^{9.5}$~cm$^{-3}$ for 
the parameter ranges explored, while it can be produced in a wide 
parameter range for the starburst metallicity; (2) a large \ion{N}{3}]/\ion{C}{3}] 
ratio for C4 can only be reproduced at a large starburst metallicity 
$Z>12Z$\sun.  At small column densities ($\log N_H<22$), the line 
ratio is high at a relative high ionization parameters; while at 
large column density, the line ratio is essentially independent of 
ionization parameters; (3) for C4, in the column density and density 
ranges considered here, \ion{N}{3}]/\ion{N}{4}] suggests a low ionization parameter 
than \ion{C}{3}]/\ion{C}{4}.

\clearpage
\begin{deluxetable}{ccc}
\tabletypesize{\scriptsize}
\tablecaption{Center and width for each component} \label{table1}
\tablewidth{0pt}
\tablehead{
\colhead{Component} & \colhead{$v$} & \colhead{$\sigma$} \\
\colhead{} & \colhead{\kms} & \colhead{\kms} \\
}
\startdata
C1 &   -1 $\pm$ 8  & 183$\pm$7 \\ 
C2 & -357 $\pm$26 & 257$\pm$18 \\ 
C3 &  -82 $\pm$79 & 933$\pm$59 \\ 
C4 & -1646$\pm$33 & 858$\pm$17 \\ 
\enddata
\end{deluxetable}

\begin{deluxetable}{ccccc}
\tabletypesize{\scriptsize}
\tablecaption{Emission line strengths for each component$^1$}\label{table2}
\tablewidth{0pt} \tablehead{
\colhead{Line} & \colhead{C1} & \colhead{C2} & \colhead{C3} & \colhead{C4}\\
}
\startdata
H$\alpha$  & 341$\pm$59 & 597$\pm$73  & 817$\pm$124 & 829$\pm$56 \\
H$\beta$   &  55$\pm$10 &  50$\pm$12  & 160$\pm$26 & 198$\pm$19   \\
H$\gamma$  &   8$\pm$7 &   11$\pm$8   & 112$\pm$49 &  77$\pm$15 \\ \
[\ion{O}{3}]5007 &  50$\pm$9  &  57$\pm$11 & 113$\pm$23 & 532$\pm$20 \\ \
[\ion{O}{3}]4959 &  17         & 19         &  38       & 213$\pm$12  \\ \
[\ion{O}{3}]4363 & \nodata    & \nodata     &  19$\pm$14 & 134$\pm$28 \\ \
[\ion{O}{3}]1663 &  \nodata   &  \nodata    &  120$\pm$31  & 100$\pm$29  \\ \
[\ion{Ne}{3}]3869&   7$\pm$8  &  68$\pm$13  &  39$\pm$33 & 308$\pm$18 \\ \
[\ion{N}{2}]6584 & 295$\pm$20 &  52$\pm$36  & \nodata & \nodata \\ \
[\ion{O}{1}]6310 & 52$\pm$6  &  12$\pm$6   & \nodata  & \nodata \\ \
[\ion{O}{2}]3727 & 294$\pm$17  & 96$\pm$15  & \nodata  & \nodata \\ \
[\ion{S}{2}]6713 & 118$\pm$9  &  35$\pm$9   & \nodata  & \nodata \\ \
\ion{C}{4}1549 & \nodata    &  \nodata    &   0         & 1403$\pm$41\\
\ion{C}{3}]1909 & \nodata    &  \nodata    & 564$\pm$45  &  196$\pm$63\\
\ion{N}{3}]1750 & \nodata    &  \nodata    & 91$\pm$27   &  751$\pm$32\\
\ion{Si}{3}]1890 & \nodata    &  \nodata    & 473$\pm$64 & 194$\pm$40\\
\ion{Al}{3}1860 & \nodata    &  \nodata    &  63$\pm$11&  55$\pm$11\\
\ion{N}{4}]1486 & \nodata     & \nodata     & \nodata   & 230$\pm$28 \\
\enddata
\tablenotetext{1}{in units of $10^{-17}$ \ergs cm$^{-2}$}

\end{deluxetable}

\clearpage


\end{document}